\documentclass[aps,prl,twocolumn,groupedaddress,showpacs]{revtex4}
\usepackage{epsfig}
\usepackage[dvipsnames,usenames]{color}
\usepackage{color}
\usepackage{amsmath}
\usepackage{graphicx}

\emergencystretch=\maxdimen
\begin{document}
\title{Tricriticality in Crossed Ising Chains}
\author{T.~Cary, R.R.P.~Singh and R.T.~Scalettar}
\affiliation{Department of Physics, One Shields Ave., University of California, Davis, California 95616, USA}

\begin{abstract}
We explore the phase diagram of Ising spins on one-dimensional chains
which criss-cross in two perpendicular directions and which are connected by
interchain couplings. This system is of interest as a simpler, classical
analog of a quantum Hamiltonian which has been proposed as a model of
magnetic behavior in Nb$_{12}$O$_{29}$ and also, conceptually, as a
geometry which is intermediate between one and two dimensions.  Using
mean field theory as well as Metropolis Monte Carlo and Wang-Landau
simulations, we locate quantitatively the boundaries of four ordered
phases.  Each becomes an effective Ising model with unique effective
couplings at large interchain coupling.  Away from this limit we
demonstrate non-trivial critical behavior, including tricritical points
which separate first and second order phase transitions. Finally, we present
evidence that this model belongs to the 2D Ising universality class.

\end{abstract}

\pacs{71.10.Fd, 71.30.+h, 02.70.Uu}
\maketitle
\section{Introduction}
Dimensionality, along with order parameter symmetry, plays a decisive
role in the occurrence of phase transitions and the critical exponents
with which they are characterized\cite{lavis15}.  Beginning with simple,
regular geometries, critical properties are now well-understood in more
complex geometries in which the dimensionality is more ambiguous,
including diluted lattices\cite{yeomans79}, fractal
geometries\cite{gefen80}, and networks with longer range
interactions\cite{baker63,nagle70,scalettar91,gitterman00,lopes04}.

Recently there has been interest in a further class of systems of ``mixed
geometry" whose underlying structure consists of two perpendicular
collections of one dimensional chains which are then further connected
to form a two dimensional framework.  For example, it has been
suggested\cite{lee15} that an appropriate model of magnetic phase transitions
in one of the niobates, Nb$_{12}$O$_{29}$, consists of one dimensional
chains of localized (Heisenberg) spins and a further perpendicularly
oriented set of one dimensional conduction electron chains.  These two
types of spins reflect the presence of distinct Nb cations with 4d$^1$
configuration, one of which exhibits local moment behavior and the other
being itinerant and Pauli paramagnetic\cite{cava91,andersen05}.  In this
model, the electron spins on the conducting nanowires are coupled to the
Heisenberg chains by a Kondo interaction on each site.

Similarly, in optical lattices\cite{greiner08}, bosonic or fermionic
atoms can occupy higher, spatially anisotropic, $p_x$ and $p_y$ orbitals
which allow hopping which is essentially just along one-dimensional
chains.  Within a given well, atoms can convert from occupying the $p_x$
to occupying the $p_y$ orbital, thus coupling the perpendiular chains and
providing a two dimensional character to the system.  Bosonic systems in
this geometry can exhibit exotic forms of superfluidity whose condensate
wave functions belong to non-trivial representations of the lattice
point group, with condensation accompanied by unusual columnar,
antiferromagnetic, and Mott phases \cite{isacsson05,liu06,wu09,hebert13}.  
Models in which fermionic
degrees of freedom in the two orbitals have Hund's rule type coupling
have also been considered, and shown rigorously to exhibit magnetic order\cite{li14}.

These examples share a common ``${\rm 1D} \times {\rm 1D}$" geometrical
structure in which one type of chain has degrees of freedom which are
coupled in the $\hat{x}$ direction, while the degrees of freedom of the
other couple in the $\hat{y}$ direction. An additional interaction on
each lattice site connects the two sets of chains. Although considerable
progress has been made in modeling the niobates and p-wave bosons in
optical lattices, in both cases the quantum nature of the spins makes
achieving a definitive understanding of the critical phenomena quite
challenging. The goal of this paper is to examine a classical Ising
model on this type of lattice. We will show that the interchain coupling
is sufficient to promote long range order at finite temperature, and
that the phase transitions can exhibit a rich variety of behaviors
including tricritical points. 

\section{Model and Methods}

We consider the following model,
\begin{align}
E = &-J_x \sum_{\bf r} S_{\bf r} S_{{\bf r}+\hat x}
  -J_y \sum_{\bf r} T_{\bf r} T_{{\bf r}+\hat y}
\nonumber \\
   &-J_z \sum_{\bf r} S_{\bf r} T_{{\bf r}}-J_{z^\prime} \sum_{\bf r} S_{\bf r} 
\, ( \, T_{{\bf r}+\hat y} + T_{{\bf r}-\hat y} \, ) 
\nonumber \\
  &-J_{z^\prime} \sum_{\bf r} T_{\bf r} 
\, ( \, S_{{\bf r}+\hat x} + S_{{\bf r}-\hat x} \, ) 
\label{eq:model}
\end{align}
which we will refer to as the crossed Ising chains model (CICM).

\begin{figure}[h!]
\includegraphics[height=6.75cm,width=9.0cm]{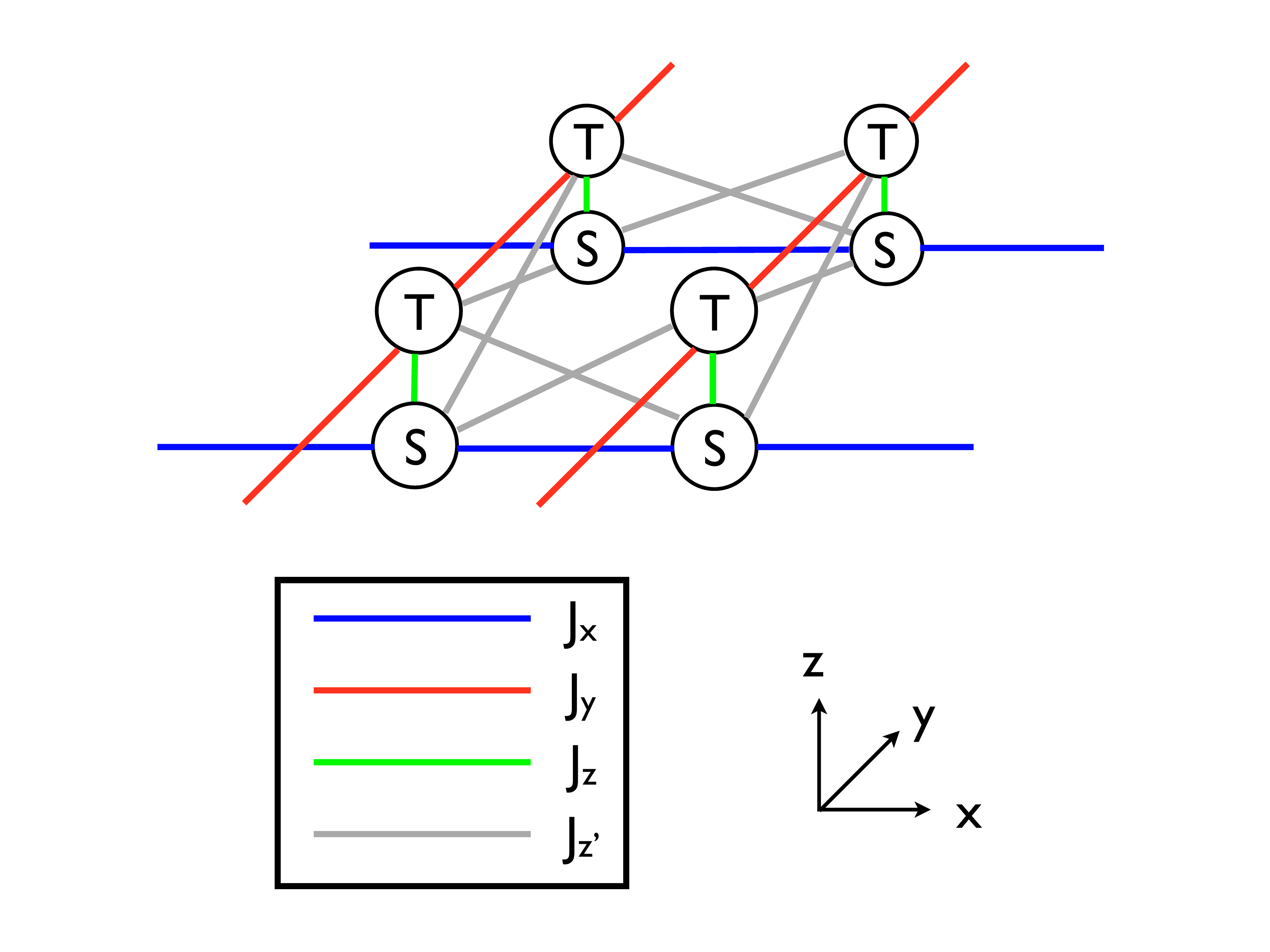}
\caption{(Color online) The geometry of the interactions in the CICM for
four sites is shown. This model is studied on an $L$ $\times$ $L$ square lattice
with periodic boundary conditions (in the $\hat x$ and $\hat y$
directions) containing $N$ (= $L^{2}$) ``S spins" and N ``T spins" taking
values $\pm 1$.
There are $L$ parallel 1D chains of S spins in the $\hat x$ direction and
$L$ parallel 1D chains of T spins in the $\hat y$ direction, illustrated
by the blue ($J_x$) and red ($J_y$) lines, respectively. There is an
interaction between an S and a T spin on the same site in the $\hat z$
direction ($J_{z}$) illustrated by the green lines. Finally, there is an
interaction between nearest neighbor S and T spins ($J_{z^\prime}$) which is
illustrated by the gray lines. }
\label{figuremodel}
\end{figure}

Here $S_{\bf r}$ and $T_{\bf r}$ are Ising spins (i.e. they can have a
value of either +1 or $-1$) coupled into one-dimensional chains in the
$\hat x$ and $\hat y$ directions, respectively.  These spins occupy a 
two-dimensional, square, $L$ $\times$ $L$ lattice with periodic boundary conditions.
There is an S and a T spin on each of the $N$=$L^2$ sites and therefore,
2$N$ total spins in the system. $J_z$ and $J_{z^\prime}$ couple S and T
spins on the same lattice site and near neighbor sites, respectively.
The geometry of Eq.~(\ref{eq:model}) is illustrated in
Fig.~\ref{figuremodel}. For
simplicity, and also because this choice is the appropriate one for
several of the physical realizations of the CICM, we will set
$J_x=J_y=J_{x,y}=1$ and measure all energies in units of $J_{x,y}$.

\begin{figure}[h!]
\includegraphics[height=6.0cm,width=8.0cm]{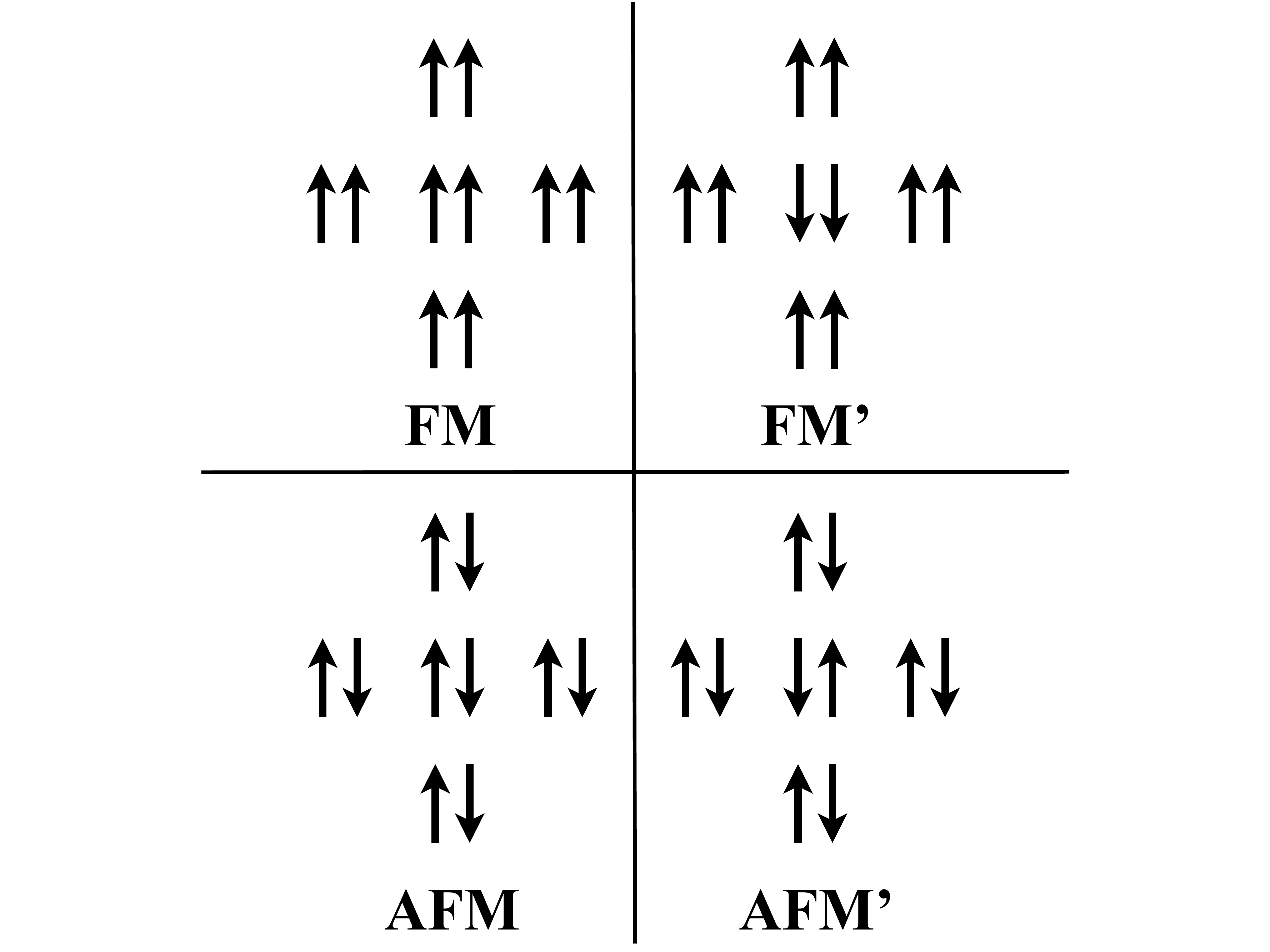}
\caption{The four ordered phases found in the CICM are defined. Each
pair of spins represents an S and a T spin on a single site (i.e.
coupled by $J_{z}$). 
In the ferromagnetic (FM) phase, all S
and T spins are aligned ferromagnetically. In the ferromagnetic prime
(FM$'$) phase, the S and T spins are aligned ferromagnetically on each
site and antiferromagnetically along the S and T chains. In the
antiferromagnetic (AFM) phase, the S and T spins are aligned
antiferromagnetically on each site and ferromagnetically along the S and
T chains. Finally, in the antiferromagnetic-prime (AFM$'$) phase, the S
and T spins are aligned antiferromagnetically on each site and also
along the S and T chains.  There is spin inversion symmetry
in this model so flipping all of the spins in any of these phases does
not change the phase.}
\label{figurephases}
\end{figure}

\begin{figure}[h!]
\includegraphics[height=8.0cm,width=8.0cm]{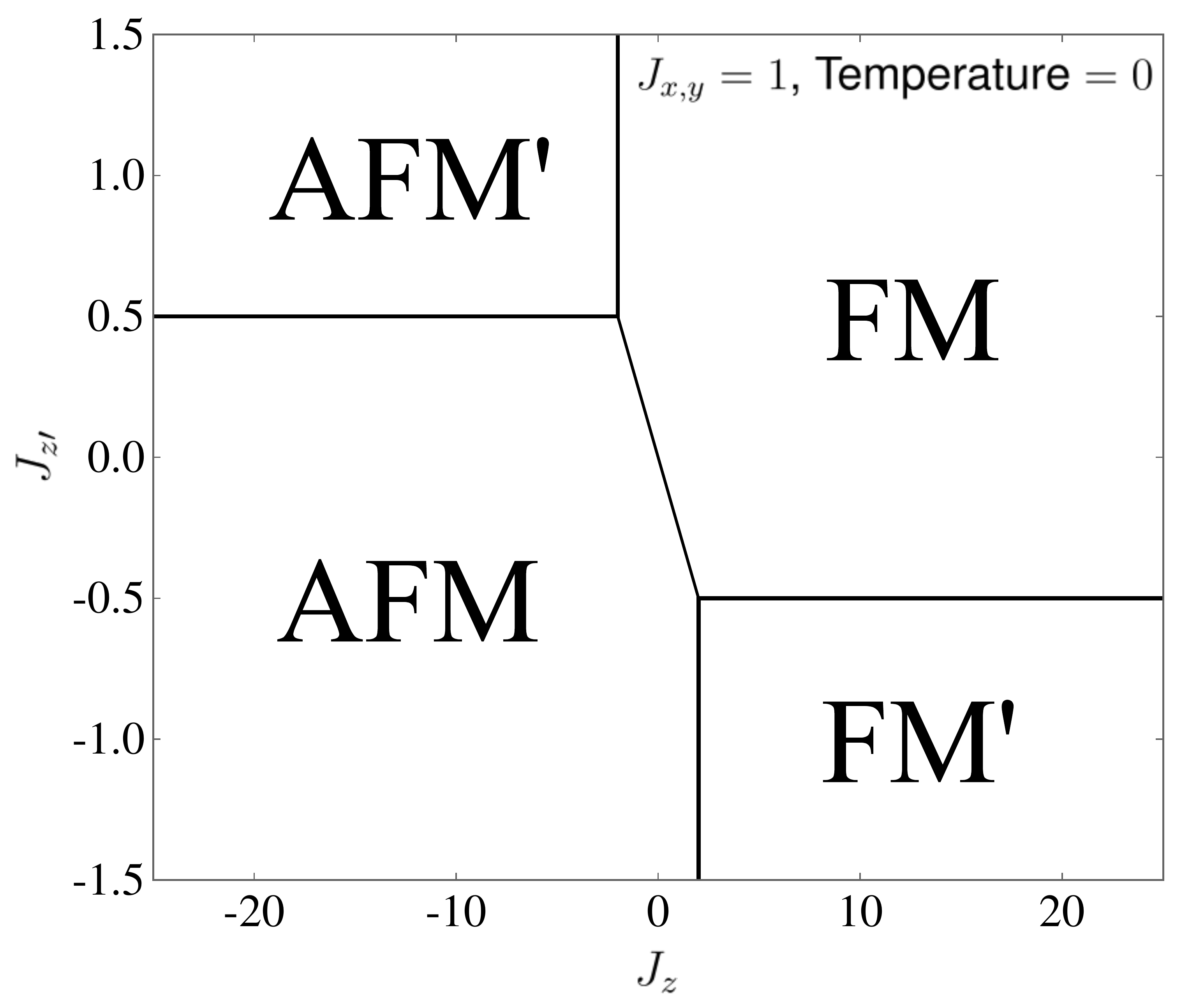}
\caption{The phase diagram for the CICM with $J_{x,y}$=1 and T=0 in the
$J_{z^\prime}$ versus $J_{z}$ parameter space is shown. The line
separating the FM and AFM phases (where the internal energies are equal)
is given by $J_{z^\prime}$=$-\frac{1}{4} J_{z}$. The AFM and AFM$'$
phases are separated by $J_{z^\prime}=\frac{J_{x,y}}{2}=\frac{1}{2}$; the
FM and FM$'$ phases 
by $J_{z^\prime}=\frac{-J_{x,y}}{2}=\frac{-1}{2}$; the FM and AFM$'$
phases  by $J_z = -2J_{x,y}=-2$; and the AFM and FM$'$ phases by
$J_z = 2J_{x,y}=2$. This phase diagram is consistent with 
the observation 
that for $J_z$ and $J_{z^\prime}$ both positive or both negative, there
is no competition between phases. Additionally, the symmetry between FM
and AFM and FM$'$ and AFM$'$ when switching the signs of $J_z$ and
$J_{z^\prime}$ is evident.}
\label{Teq0phases}
\end{figure}

Initial insight into the phase diagram of this model is obtained by
considering $T=0$ and minimizing the internal energy,
Eq.~(\ref{eq:model}).  Fig.~\ref{figurephases} 
shows the definitions of the four ordered
phases which can occur: ferromagnetic (FM), ferromagnetic-prime (FM$'$),
antiferromagnetic (AFM), and antiferrmagnetic-prime (AFM$'$).  The phase
diagram at $T=0$ is shown in Fig.~\ref{Teq0phases}.
The CICM has the symmetry that
changing $J_z \to -J_z$ and $J_{z^\prime} \to -J_{z^\prime}$ changes the phase from FM
$\to$ AFM or AFM $\to$ FM, and FM$'$ $\to$ AFM$'$ or AFM$'$ $\to$ FM$'$.  If $J_z$
and $J_{z^\prime}$ are both positive or both negative, there will be no
competition between ordered phases and the model will have relatively
uninteresting features, namely a conventional second order phase
transition between a high temperature disordered paramagnetic (PM) phase
and a low temperature FM phase or AFM phase, respectively.  However, if
only one of the interchain couplings is negative, there will be a
competition between ordered phases and the most interesting physics
will result.

The total spin, $S_{\bf r}+T_{\bf r}$, on a site can take on the three
values, $-2,0,\text{or}+2$, giving the CICM some similarity to the
two-dimensional square lattice Blume-Capel model
\cite{blume66,capel66},
\begin{equation} 
E = - J \sum_{\langle ij\rangle} M_{i} M_{j} + \Delta \sum_{i} M_{i}^{2}
\end{equation}
which is a spin 1 generalization of the Ising model where
$M_{i}=-1,0,\text{or}+1$.  The choice $J_z<0$ favors $S_{\bf r}=-T_{\bf r}$
and hence $S_{\bf r}+T_{\bf r}=0$ so that the strength of $J_z$
plays a role similar to that of the vacancy potential $\Delta$ whose
energy $\Delta \, M_{i}^2$ can tune the density of sites with $M_{i}=0$.

The remainder of this paper is organized as follows.  We begin our
discussion of Eq.~(\ref{eq:model}) via a mean field treatment. The
resulting phase diagrams, as in the case of the BCM, will be shown to
correctly predict certain qualitative features of the CICM such as the
presence of ordered phases, effective Ising
regimes in the large $|J_{z}|$ limit, and
tricritical points. We then turn to a Monte Carlo (MC) approach
which allows a more accurate quantitative determination of the phase
diagram. We use the standard single-spin flip Metropolis MC algorithm,
supplemented by some multiple-spin flips. 
The data are analyzed with standard
numerical approaches, including the use of the Binder fourth order
cumulant \cite{landau00}.  The results show that there are four ordered
phases, each of which becomes an effective Ising model in the large
$|J_{z}|$ limit with unique effective couplings.
Additionally, the presence of tricritical points is confirmed. 
In order to provide further corroboration for the nature of the phase
transitions, we also employ the Wang-Landau
algorithm\cite{wang01a,wang01b,landau02} to obtain the density of states
and the behavior of canonical distributions as a function of temperature
when passing through first and second order phase transitions. We find
that this algorithm is particularly well suited for verifying the order
of a phase transition and therefore the existence of tricritical points.
Finally, we use finite-size scaling techniques to verify the universality 
class of the CICM. 

\section{Mean Field Theory}
We solve Eq.~(\ref{eq:model}) by
replacing the two spin interactions with a single spin coupled to a
self-consistently determined average spin value

\begin{equation}
m_{1} \equiv <S_{r}> \hskip1.0in m_{2} \equiv <T_{r}>.
\end{equation}

In the case of the FM' and AFM' phases, these order parameters alternate
in sign on the (bipartite) lattice.

The resulting implicit equations for the order parameters,
$m_{FM}$=$m_{1}$=$m_{2}$ and $m_{AFM}$=$m_{1}$=$-m_{2}$
($\beta = \frac{1}{T}$ and $k_{B}=1$),   
\begin{align}
m_{FM} &= \frac{\sinh{(4\beta m_{FM} (J_{x,y} + 2J_{z^\prime} ))}}
{\cosh{(4\beta m_{FM} (J_{x,y} + 2J_{z^\prime} ))}+e^{-2 \beta J_{z}}}
\nonumber \\
m_{AFM} & = \frac{\sinh{(4\beta m_{AFM} (J_{x,y} - 2J_{z^\prime} ))}}
{\cosh{(4\beta m_{AFM} (J_{x,y} - 2J_{z^\prime} ))}+e^{2 \beta J_{z}} }
\end{align}
are solved using Newton's method.  Equivalently, the mean field free
energy of the CICM can be expanded in a power series in the order
parameter for both the FM and AFM phases and the critical temperature
for a second order phase transition determined by calculating the
temperature where the coefficient of the quadratic term in the free
energy expansion vanishes. The implicit equations for the FM and AFM
second order critical lines are as follows.

\begin{align}
T_{\text{C,FM}}&=\frac{4(J_{x,y}+2J_{z^\prime})}{1+e^{\frac{-2J_{z}}{T_{\text{C,FM}}}}}
\nonumber \\
T_{\text{C,AFM}}&=\frac{4(J_{x,y}-2J_{z^\prime})}{1+e^{\frac{2J_{z}}{T_{\text{C,AFM}}}}} .
\label{2ndorder}
\end{align}

Tricritical points are located by calculating the temperature at which
the quartic coefficient in the expansion of the free energy vanishes.
\begin{align}
T_{\text{tricritical, FM}} &= \frac{-2}{\ln(2)}J_{z}
\nonumber \\
T_{\text{tricritical, AFM}} &= \frac{2}{\ln(2)}J_{z} .
\label{tricrit1}
\end{align}
Combining this with the condition for intesecting the second order phase
boundary, simple analytic expressions for the coordinates of the mean
field tricritical points can be written down.  

\begin{align}
T_{\text{tricritical point, FM}} &= \frac{4(J_{x,y}+2J_{z^\prime})}{3}
\nonumber\\
J_{\text{z,tricritical point, FM}} &=
\frac{-2\ln(2)(J_{x,y}+2J_{z^\prime})}{3}
\nonumber\\
T_{\text{tricritical point, AFM}} &= \frac{4(J_{x,y}-2J_{z^\prime})}{3}
\nonumber\\
J_{\text{z,tricritical point, AFM}} &=
\frac{2\ln(2)(J_{x,y}-2J_{z^\prime})}{3}
\label{tricrit2}
\end{align}

\begin{figure}[h!]
\includegraphics[height=9.0cm,width=9.0cm]{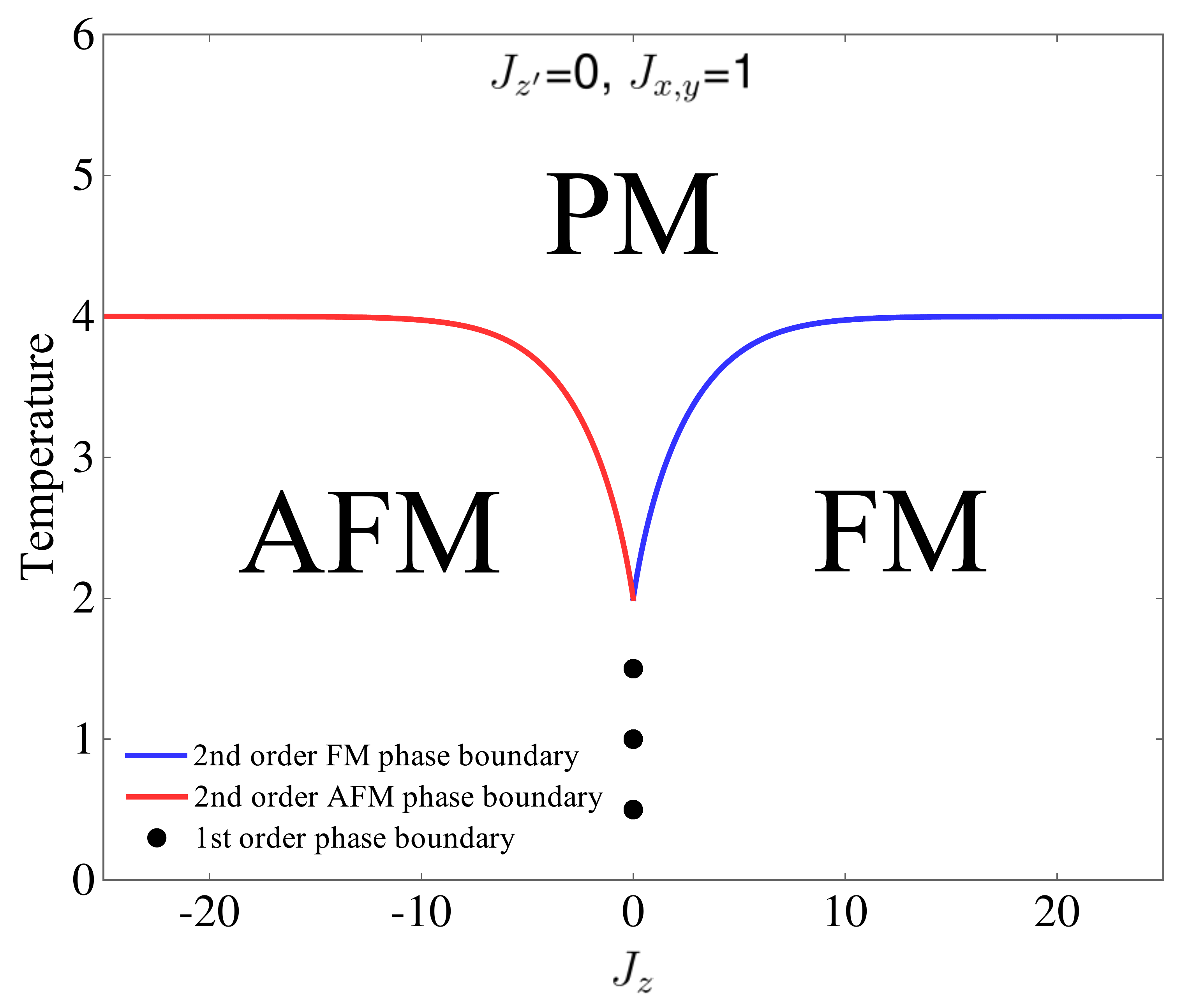}
\caption{(Color online) The mean field theory phase diagram for
$J_{z^\prime}=0$ and $J_{x,y}=1$ is shown. 
For these parameters, there is no tricritical point. For $J_z >$ 0, there is a
second order phase transition between a low temperature FM phase and a
high temperature PM phase. For $J_z$ $<$ 0, there is a second order
phase transition between a low temperature AFM phase and a high
temperature PM phase. Also, there is a vertical first order
phase boundary between the FM and AFM phases at $J_z$= 0 which extends
up to T=2, the MF critical temperature for the 1D Ising model.
In the large, positive $J_z$ limit, this model
becomes an effective 2D Ising model with
$T_C = 4 (J_{x,y}+2J_{z^\prime})=4$. In the large,
negative $J_z$ limit, this model also
becomes an effective 2D Ising model with
$T_C = 4 (J_{x,y}-2J_{z^\prime})=4$. 
The FM and AFM phase shapes are symmetric about $J_z$=0 only when $J_{z'}$=0.}
\label{mftJkp00}
\end{figure}

\begin{figure}[h!]
\includegraphics[height=9.0cm,width=9.0cm]{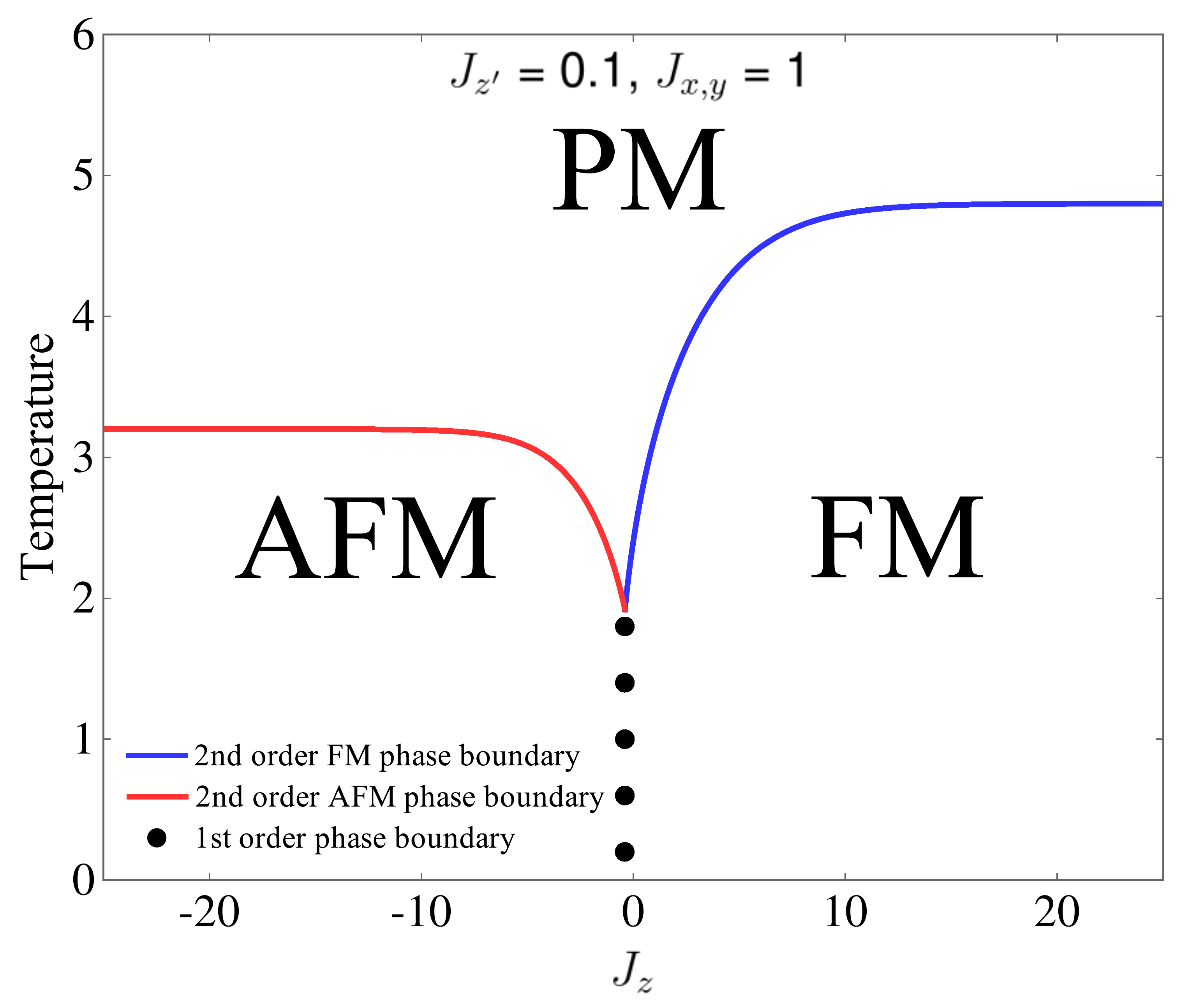}
\caption{(Color online) The mean field theory phase diagram for
$J_{z^\prime}=0.1$ and $J_{x,y}=1$ is shown. 
For these parameters, there is no tricritical point. For $J_z > -4 J_{z^\prime}
= -0.4$, there is a
second order phase transition between a low temperature FM phase and a
high temperature PM phase. For $J_z$ $< -4 J_{z^\prime} = -0.4$, there is a second order
phase transition between a low temperature AFM phase and a high
temperature PM phase. Also, there is a vertical first order
phase boundary between the FM and AFM phases at $J_z= -0.4$.
In the large, positive $J_z$ limit, this model
becomes an effective 2D Ising model with
$T_C = 4 (J_{x,y}+2J_{z^\prime})=4.8$. In the large,
negative $J_z$ limit, this model also
becomes an effective 2D Ising model with
$T_C = 4 (J_{x,y}-2J_{z^\prime})=3.2$. 
A positive 
$J_{z^\prime}$ shrinks the AFM phase and grows the FM phase as it is increased
in magnitude, as evidenced by comparing with the phase diagram for 
$J_{z^\prime}=0$. }
\label{mftJkp01}
\end{figure}

\begin{figure}[h!]
\includegraphics[height=9.0cm,width=9.0cm]{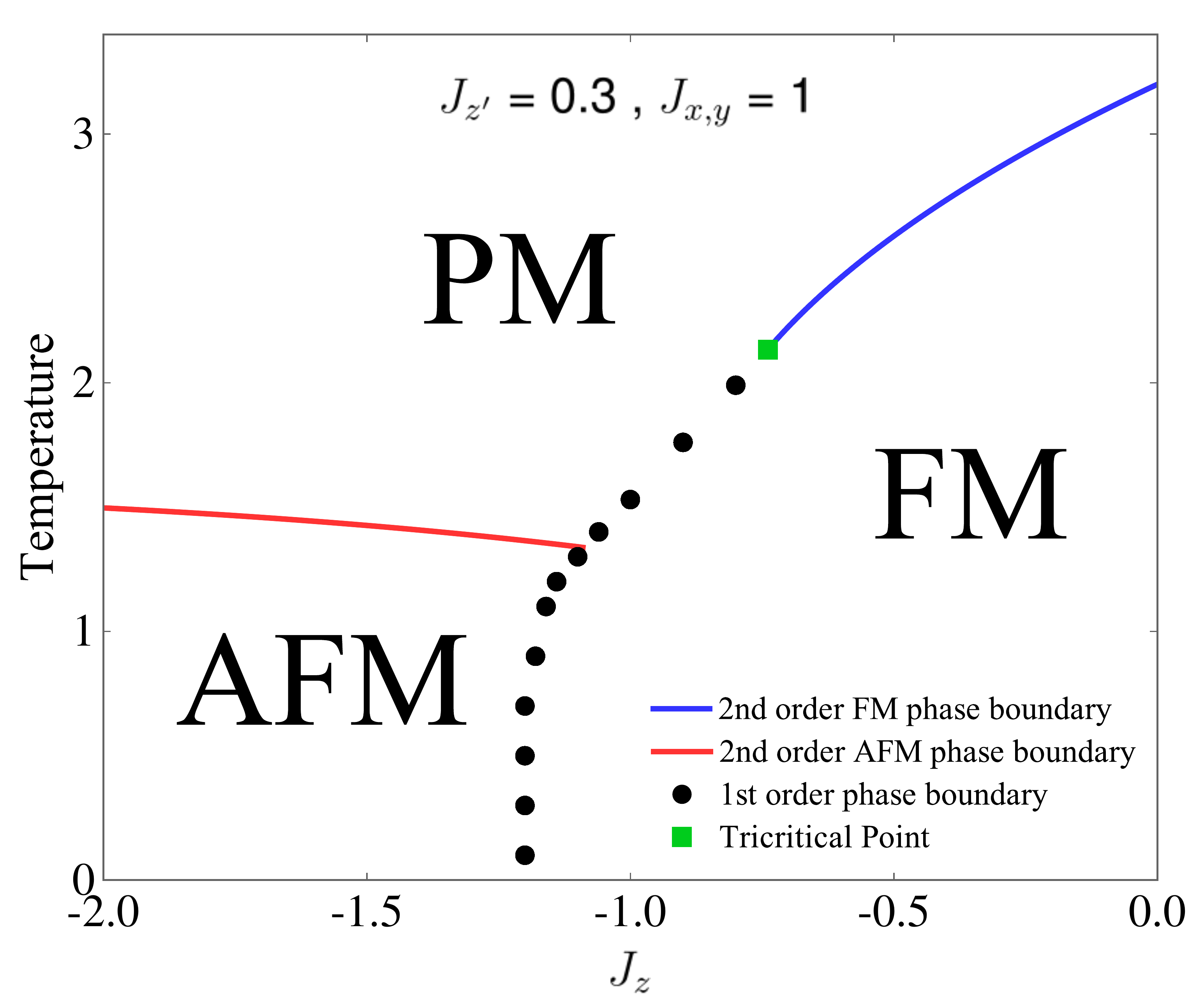}
\caption{(Color online) The mean field theory phase diagram for
$J_{z^\prime}=0.3$ and $J_{x,y}=1$ is shown. For $J_z$ $\gtrsim -0.7$, there
is a second order phase transition between a low temperature FM phase
and a high temperature PM phase. For $-1.1 \lesssim$ $J_z$ $\lesssim
-0.7$, there is a first order phase transition between a low temperature
FM phase and a high temperature PM phase. This section of the phase
diagram is separated from the previous section by the green square tricritical
point (located at T = $\frac{32}{15} \approx 2.13$, $J_z$ = $\frac{-16
\ln(2)}{15} \approx -0.74$). For $-1.2 \lesssim$ $J_z$ $\lesssim -1.1$
there is a low temperature FM phase followed by a small higher
temperature AFM phase and then, for higher temperatures, a disordered PM
phase. For $J_z$ $\lesssim -1.2$ there is a second order phase boundary
between a low temperature AFM phase and a high temperature PM phase.
Additionally, there is an approximately vertical first order phase
boundary between the FM and AFM phases at $J_z$ $=$ $-1.2$. This phase
diagram is zoomed in relative to the other phase diagrams in order to
show the details of the tricritical point and first order phase
boundary.}
\label{mftJkp03}
\end{figure}

To find the first order phase boundary once the tricritical point has
been reached, simultaneous plots of the FM and AFM free energy were made
and temperature or $J_{z}$ was incremented to find the point where the
phase with the global minimum changes. 

\begin{figure}[h!]
\includegraphics[height=9.0cm,width=9.0cm]{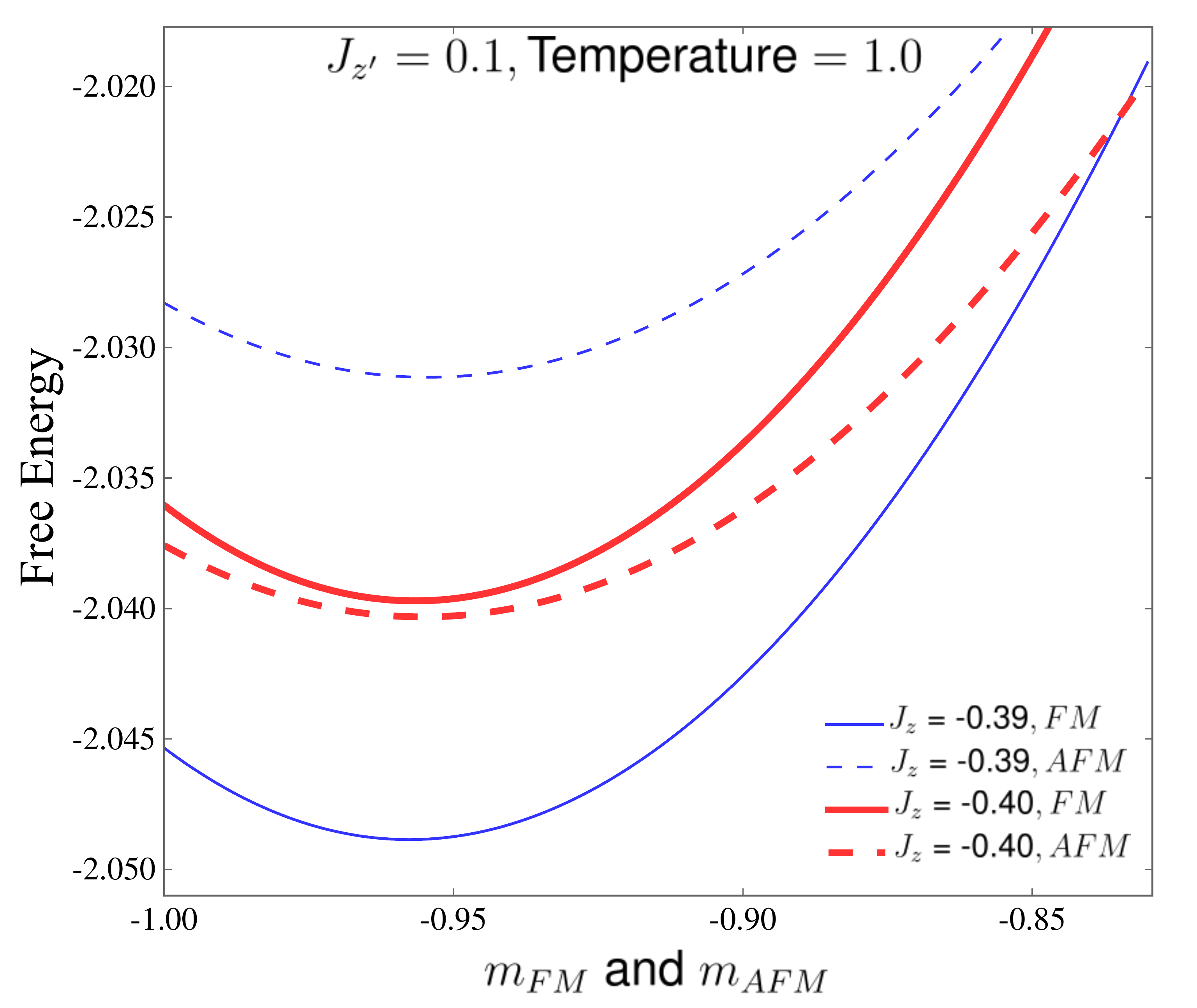}
\caption{(Color online) A plot of the FM and AFM free energies at
$J_{z^\prime}=0.1$ and T=1.0 is shown. For $J_z$=$-0.39$ the minimum of the FM
free energy (thin blue solid curve) is lower than the minimum of the AFM free
energy (thin blue dashed curve) and therefore, the system is
in the FM phase. For $J_z$=$-0.40$ the minimum of the FM free energy (thick red
solid curve) is greater than the minimum of AFM free energy (thick red dashed
curve) and therefore the system is in the AFM phase. If
the global minima were to be at m=0, the
system is in the PM phase where the total magnetization is 0. This is
the analysis used to determine all of the first order phase boundaries in
the mean field theory phase diagrams.}
\label{mftJkp01F}
\end{figure}

For $J_{z^\prime}$ = 0, the mean field phase diagram
(Fig.~\ref{mftJkp00}) shows no tricritical point. Clearly,
$J_{z^\prime}$ is necessary for the onset of first order phase
transitions. 
The AFM phase arises for $J_z < 0$, as expected, since negative $J_z$
antiferromagnetically couples the $S$ and $T$ spins on shared sites. For
$J_z > 0$, the FM phase arises. The mean field phase boundary separating
the FM and AFM phases at $T=0$ where thermal fluctuations are
nonexistent agrees with the ground state phase diagram in
Fig.~\ref{Teq0phases}.
The MFT critical temperature is $T_C=2$ at $J_{z}$=0 and $J_{z^\prime}$=0, as
expected since the CICM decouples into independent 1D Ising chains
For large $|J_z|$, the $S$ and $T$ spin pairs on each site lose
their independence due to the high energy cost of
flipping only one of the spins in a pair. In this limit, the model
becomes an effective 2D Ising model with 
$J_{\text{eff,FM}}=J_{x,y}+2J_{z^\prime}$ and
$J_{\text{eff},AFM}=J_{x,y}-2J_{z^\prime}$ for positive and
negative $J_z$, respectively. 
This leads to the limiting values $T_C=4$ for
$|J_{z}|$ large in Fig.~\ref{mftJkp00}. 

In fact, this single ``locked spin"
Ising regime in the large $|J_{z}|$ limit
occurs for all four ordered phases. However, the effective couplings are
different for each phase. In the large, negative $J_{z}$ limit, the AFM
and AFM$'$ phases have the following effective Ising couplings.
\begin{align}
J_{\text{eff,AFM}}&=J_{x,y}-2J_{z^\prime}
\nonumber \\
J_{\text{eff,AFM$'$}}&=-J_{x,y}+2J_{z^\prime}
\end{align}
Meanwhile, in the large, positive $J_{z}$ limit, the FM and FM$'$ phases 
have the following different effective Ising couplings.
\begin{align}
J_{\text{eff,FM}}&=J_{x,y}+2J_{z^\prime}
\nonumber \\
J_{\text{eff,FM$'$}}&=-J_{x,y}-2J_{z^\prime}.
\end{align}
This behavior is similar to that of the Blume-Capel
Model (BCM) which also approaches an Ising limit
for large negative $\Delta$ which drives the density of vacancy
sites $M_i=0$ to zero. However, our model does not approach 
the ``vacant" lattice limit of the BCM at large positive 
$\Delta$, because even though $S_i=-T_i$ in the AFM and AFM$^\prime$
phases, the individual nonzero $S$ and $T$ moments still couple down
their respective chains. It is interesting, therefore, that, as we shall see, the tricritical
points which are driven by vacancies in the BCM are still present in the CICM.

When $J_{z^\prime} \ne 0$  the phase diagram loses its symmetry on
changing the sign of $J_{z}$. 
As expected, for $J_{z^\prime}$=0.1 (Fig.~\ref{mftJkp01}), the AFM and FM
phases meet at $J_{z}$=-4$J_{z^\prime}$=-0.4. Also, for large, negative
$J_{z}$, $T_c$ = 4($J_{x,y} - 2 J_{z^\prime}$)=3.2 and for large, postive
$J_{z}$, $T_c$ = 4($J_{x,y} + 2 J_{z^\prime}$)=4.8;  as
$J_{z^\prime}>0$ gets larger, the FM phase gets larger and the AFM phase shrinks. 
The phase diagram is reflected about $J_z$=0 for 
$J_{z^\prime}$=$-0.1$ (not shown): the AFM region expands and the FM
region shrinks.

Most importantly, the value of $J_{z^\prime}$ determines whether or not
there is a triciritcal point. For $J_{z^\prime} = 0.0$ and $0.1$, there is no
tricritical point and all thermally driven phase transitions between an
ordered phase and the disordered phase are of second order. However, for
$J_{z^\prime}$ = 0.3 (Fig.~\ref{mftJkp03})
there is a tricritical point.
The thermally driven phase transition between the PM and the FM phase
switches from second order to first order.
The FM tricritical point emerges when $J_{z^\prime}$ $>$
$\frac{J_{x,y}}{6}$=$\frac{1}{6}$, a result which follows from a detailed
analysis of Eq.~(\ref{2ndorder}) and Eq.~(\ref{tricrit2}).

\section{Metropolis Monte Carlo}

In order to achieve more accurate quantitative results, the Metropolis
MC algorithm was implemented. We include moves which 
flip a single S spin, a single T spin, a row
of S spins, a column of T spins, and an S and T spin 
simultaneously on a single site.
What we will call one sweep alternates between the following five procedures:
flipping every S spin ($N$ total flips), flipping every T spin ($N$ total flips), 
flipping every row of S spins ($L$ total flips), flipping every column of T spins
($L$ total flips), and flipping every S and T pair ($N$ total flips).
To thermalize the lattice we perform $5 \times 10^{5}$ such sweeps of the lattice
(i.e. $10^{5}$ sweeps of each type). We then perform another $5 \times 10^{5}$ sweeps 
of the lattice, making a measurement every 10 sweeps.
Flipping multiple spins at a time helps the system to break out 
of metastable states and thereby makes the algorithm more efficient. 
For example, if $J_z$ is large and positive and a pair of S and T spins 
both have values of +1, the probability of both spins changing to $-1$ 
is very small if only single spin flips are allowed. This is because 
of the large increase in energy that would come from trying to change 
the value of one of them first (i.e. making them align antiferromagnetically).

In order to calculate the critical temperatures, the Binder fourth order 
cumulant,
\begin{equation}
U_{L} = 1 - \frac{<m^4>}{3<m^2>^2}
\end{equation}
where m is either $m_{\rm FM}$, $m_{\rm AFM}$, $m_{\rm FM'}$, 
or $m_{\rm AFM'}$ is calculated as a function of
temperature for various lattice sizes, $L$.  Curves for different
lattice sizes have a common intersection point at the
critical temperature ($T_C$), regardless of the order of the transition \cite{vollmayr93}. 
Additionally, the behavior of the Binder cumulant 
away from the crossing at $T_C$ can be used to
distinguish between first and second order phase transitions. For second
order phase transitions, $U_{L}$ approaches the value 
$U_{L}=\frac{2}{3}$ as the temperature approaches zero. For temperatures above
the critical temperature, $U_{L}$ approaches $U_{L}=0$, all the while
staying between these two values. For first order phase transitions, the
Binder cumulant has the same limit values but, above the transition
temperature, it develops a minimum that dips below 0 and which gets deeper
for larger lattice sizes \cite{vollmayr93}.  

\begin{figure}[h!]
\includegraphics[height=9.0cm,width=9.0cm]{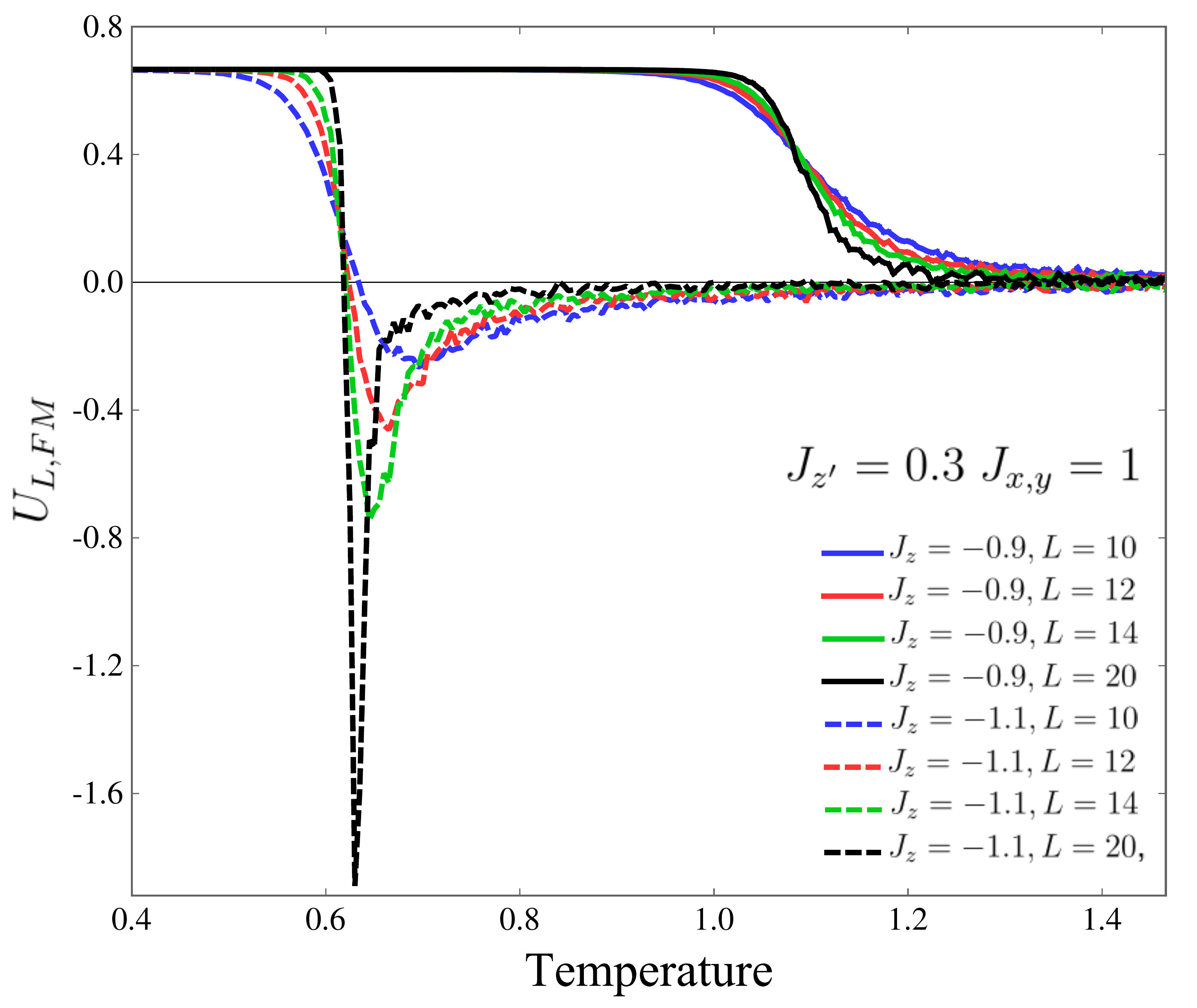}
\caption{(Color online) The Binder fourth order cumulants for various
lattice sizes and for a second order phase transition (solid lines) at
$J_z=-0.9$ and a first order phase transition (dashed lines) at
$J_z=-1.1$. A minimum below $U_{L,{\rm FM}}$=0, which gets deeper as
the lattice size increases, is a signature of a first order phase
transition.  The curves intersect
at a single critical temperature in both cases. In both cases,
at lower temperatures than the crossing, the order of the curves from top to bottom are
$L$=20, $L$=14, $L$=12, and finally $L$=10. $U_{L}$  was used to
determine all of the critical points in the MC phase diagrams.}
\end{figure}

\begin{figure}[h!]
\includegraphics[height=9.0cm,width=9.0cm]{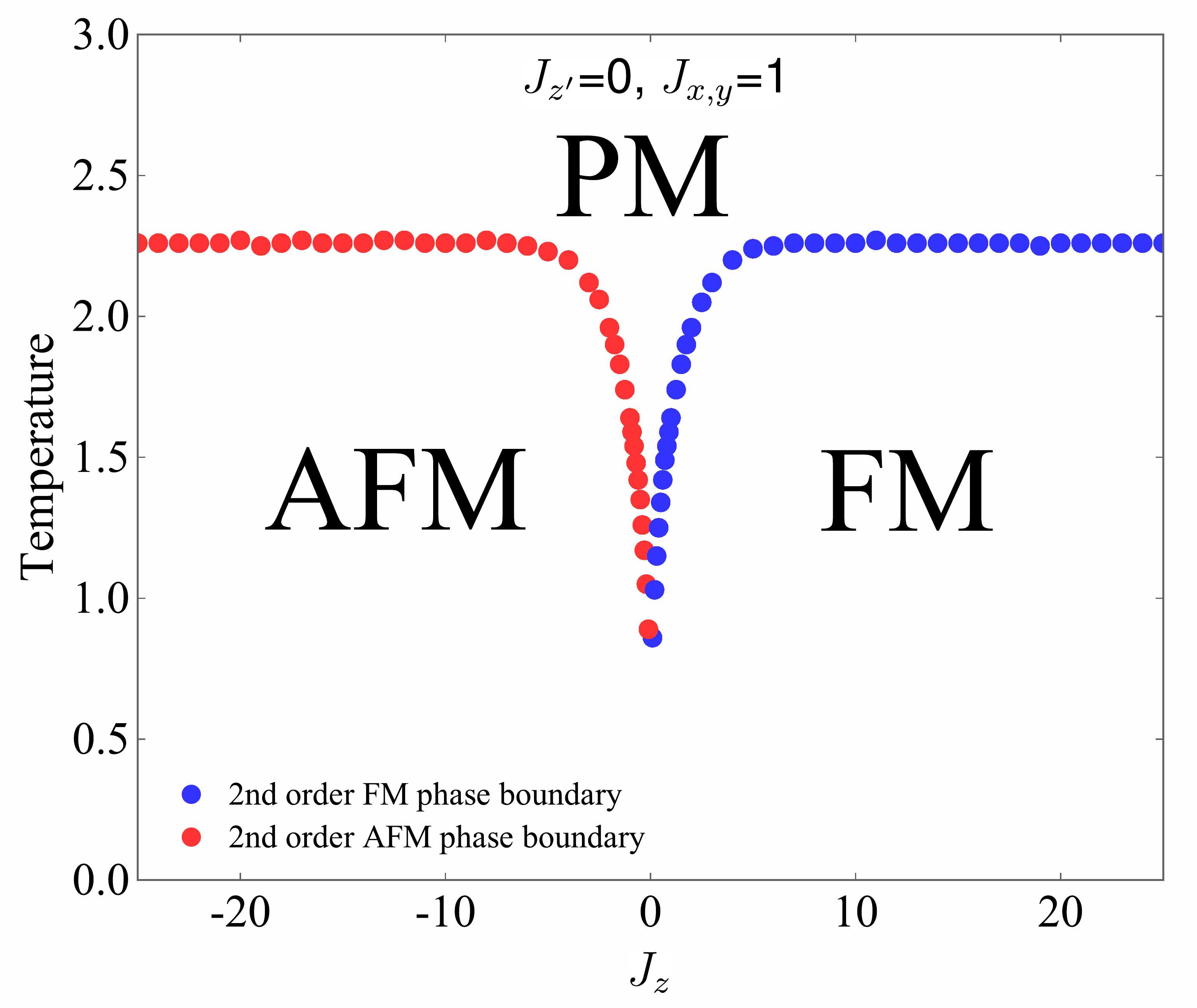}
\caption{(Color online) The MC-derived phase diagram for
$J_{z^\prime}=0$ and $J_{x,y}=1$ is shown. For $J_z > $ 0, there is a
second order phase boundary separating a low temperature FM phase from a
high temperature PM phase. The critical temperature increases as $J_z$
increases and saturates at
$T_{C} \approx 2.269 (J_{x,y}+2J_{z^\prime}) = 2.269$.
For $J_z < $ 0, there is a second order phase boundary
separating a low temperature AFM phase from a high temperature PM phase.
Similarly, the critical temperature increases as $J_z$ decreases until it saturates at 
with $T_{C} \approx 2.269 (J_{x,y}-2J_{z^\prime}) = 2.269$. This
phase diagram is qualitatively similar to the MFT phase diagram with the
same parameters, particularly in its lack of a tricritical point. 
As expected, the critical temperatures were overestimated in MFT.}
\label{MCphasediagramJKp00}
\end{figure}

\begin{figure}[h!]
\includegraphics[height=9.0cm,width=9.0cm]{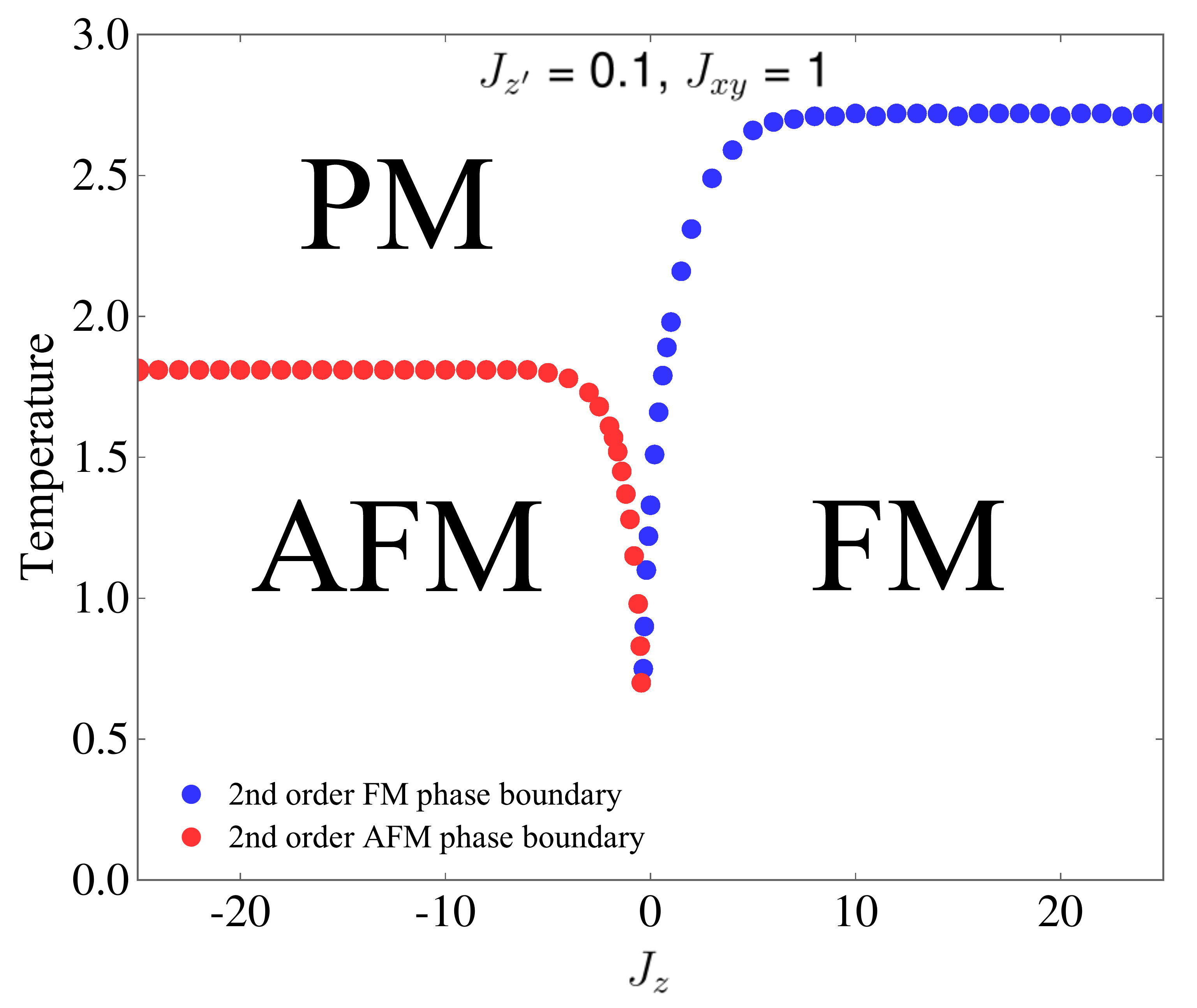}
\caption{(Color online) 
Same as Fig.~\ref{MCphasediagramJKp00} except $J_{z^\prime}=0.1$.
The phase diagram is qualitatively similar to the result of MFT,
particularly in its lack of a tricritical point.
}
\label{MCphasediagramJKp01}
\end{figure}

\begin{figure}[h!]
\includegraphics[height=9.0cm,width=9.0cm]{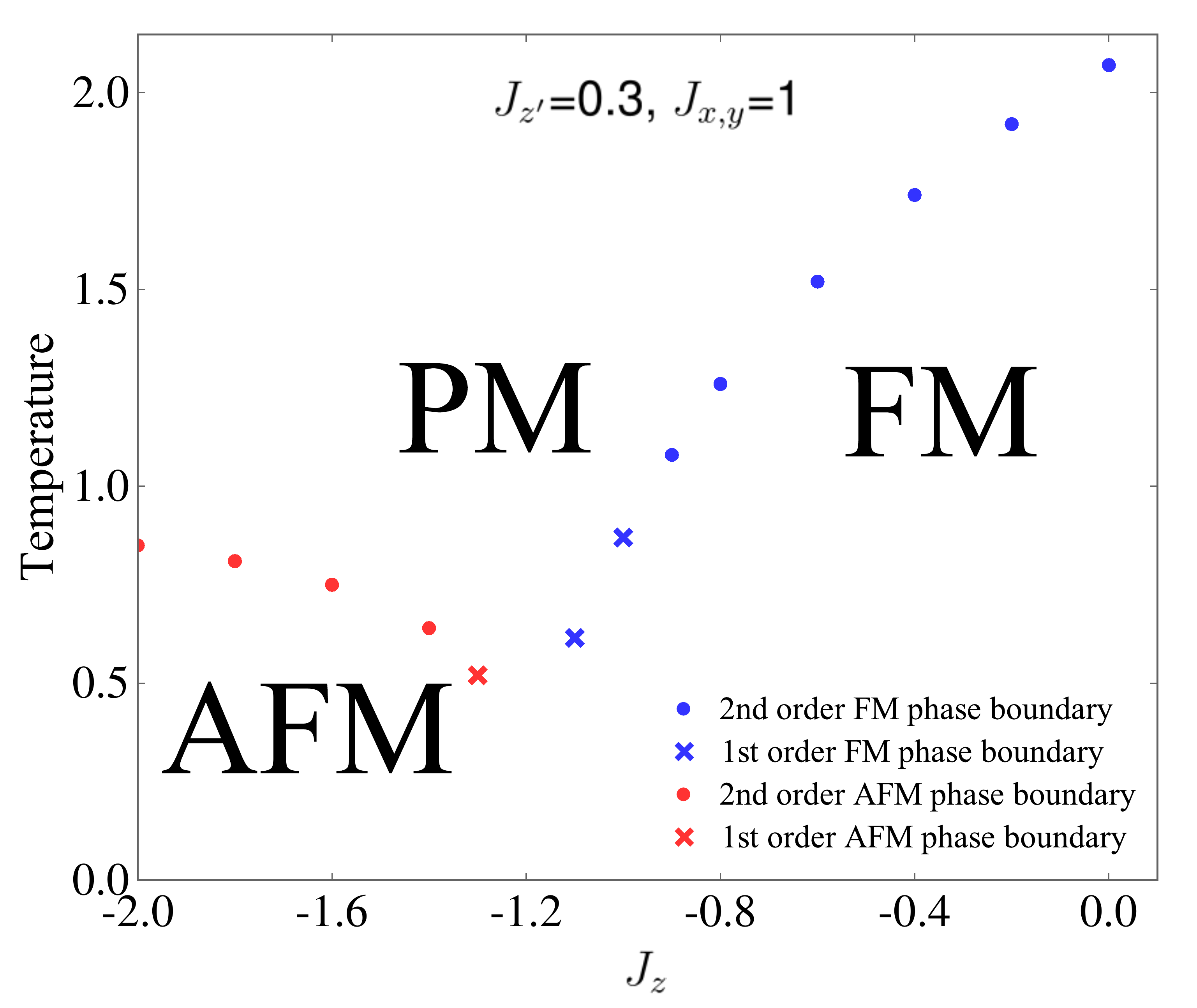}
\caption{(Color online) 
Same as Fig.~\ref{MCphasediagramJKp00} except $J_{z^\prime}=0.3.$
The phase diagram is qualitatively similar to the result of MFT in most 
regards.  However, MC finds that a tricritical point, which is present only
on the FM side in MFT, is also present on the AFM phase boundary.}
\label{MCphasediagramJKp03}
\end{figure}

\begin{figure}[h!]
\includegraphics[height=9.0cm,width=9.0cm]{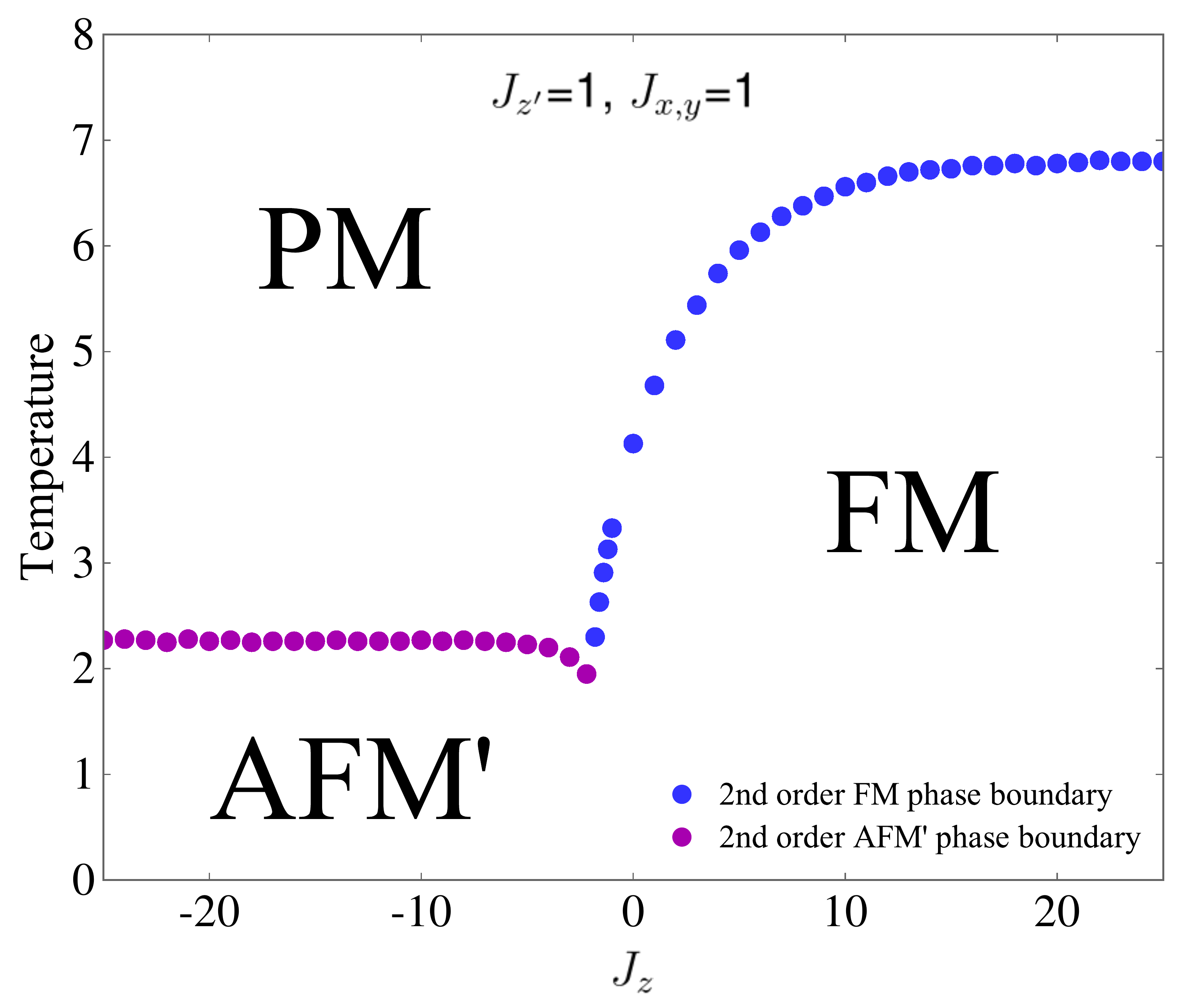}
\caption{(Color online) 
Same as Fig.~\ref{MCphasediagramJKp00} except $J_{z^\prime}=1.0.$
As for $J_{z^\prime}=0.0$ and $0.1$, there is no
tricritical point, which thus 
appear to be restricted to intermediate $J_{z^\prime}\approx 0.3$.}
\label{MCphasediagramJKp10}
\end{figure}

For $J_{z^\prime}$=0, the MC phase diagram 
of Fig.~\ref{MCphasediagramJKp00}
has the same qualitative
features as the mean field phase diagram. In both cases there are two
ordered phases at low temperatures, FM and AFM, and a PM phase at high
temperatures. The AFM and FM phases meet, as expected, at
$J_z$=$-4 J_{z^\prime}=$0. Additionally, the MC phase diagram also
contains the expected Ising regimes at large $|J_{z}|$,
that is, $T_C \approx 2.269 J_{\text{eff}}$ \cite{onsager44}.
For large positive $J_{z}$ this leads to $T_C$ $\approx$ 2.269
($J_{x,y}+2J_{z^\prime})=2.269$ and for large negative $J_{z}$, 
$T_C$ $\approx$ 2.269 ($J_{x,y}-2J_{z^\prime})=2.269$. We can estimate 
the error bars on the MC simulations by comparing how close the 
MC data is to the exact value in the Ising regime. This leads to
error bars on the critical temperatures of $\pm 0.02$.  Another way
of quantifying the uncertainty in the values of the critical temperatures 
is to estimate the ``spread" in the crossings of the fourth order Binder cumulants
for the various lattice sizes,
since the crossings are not perfectly sharp. This measure also leads to
error bars on the critical temperatures of $\pm 0.02$.

Similarly, for $J_{z^\prime}$=0.1 
(Fig.~\ref{MCphasediagramJKp01})
the MC phase diagram qualitatively
agrees with the mean field theory phase diagram. There is no tricritical
point in either the mean field theory or MC phase diagrams and the FM
and AFM phases meet at $J_z$=$-4 J_{z^\prime}= -0.4$ in both cases. For
large positive $J_{z}$, we expect $T_C$ $\approx$ 2.269
($J_{x,y}+2J_{z^\prime}$) $\approx$ 2.723 and for large negative
$J_{z}$, we expect $T_C$ $\approx$ 2.269 ($J_{x,y}-2J_{z^\prime}$) $\approx$ 1.815, which agrees with the MC data. 

Fig.~\ref{MCphasediagramJKp03} shows MC results for
$J_{z^\prime}$=0.3.  The FM and AFM phases meet at
$J_{z}$=-4$J_{z^\prime}= -1.2$, as in the MF phase diagram, and there is
a FM tricritical point at $J_z=-0.9(1)$. One important qualitiative difference between
the MF and MC phase diagrams for $J_{z^\prime}$=0.3 is
that there is also an AFM tricritical point in the MC phase diagram. 
The MF phase diagram also has a small parameter window
for  $-1.2 \lesssim$ $J_z$ $\lesssim -1.1$ where raising the temperature from the FM phase results in
passage through an intermediate AFM phase before the disordered high
temperature regime is reached.  We do not observe this in the MC data.

Finally, in Fig.~\ref{MCphasediagramJKp10} the phase diagram for 
$J_{z^\prime}$=1.0 is shown. The FM and AFM' phases meet at $J_z=-2$,
as expected from the ground state phase diagram. There is no tricritical
point for this value of $J_{z'}$ which shows that there is some intermediate range
between $J_{z'}$=0.1 and $J_{z'}$=1.0 where tricritical points are present.

\section{Wang-Landau Sampling}

While the Metropolis MC algorithm is the most widely used method of
numerically calculating the thermodynamic properites of classical spin
models, there exist more sophisticated alternatives. One is Wang-Landau sampling
(WLS). In WLS, the density of states (DOS) is determined using a MC procedure.
From the DOS, all of the desired thermodynamic properties can be calculated.
The major advantage of WLS
is that the DOS is independent of temperature so that only one
simulation is needed to calculate thermodyanamic quantities at any
temperature. Additionally, the DOS can be used to
calculate the unnormalized canonical distribution, $P(E)$,
\begin{equation}
P(E) \propto g(E) \, e^{-\beta E}
\end{equation}
for various temperatures from one simulation.
This distribution is another useful tool for distinguishing
between first and second order phase transitions as it has distinct
behavior in the two cases. For
second order phase transitions, the canonical distribution is always a
single peaked distribution which shifts its average value as the temperature
changes. For first order phase transitions, the canonical
distribution is similarly a single peaked distribution at temperatures well
above and below the phase boundary. However, it develops a
characteristic double peaked structure near the transition temperature due
to phase coexistence.  The peaks are of equal height at the transition
temperature \cite{challa86}. 

\begin{figure}[h!]
\includegraphics[height=9.0cm,width=9.0cm]{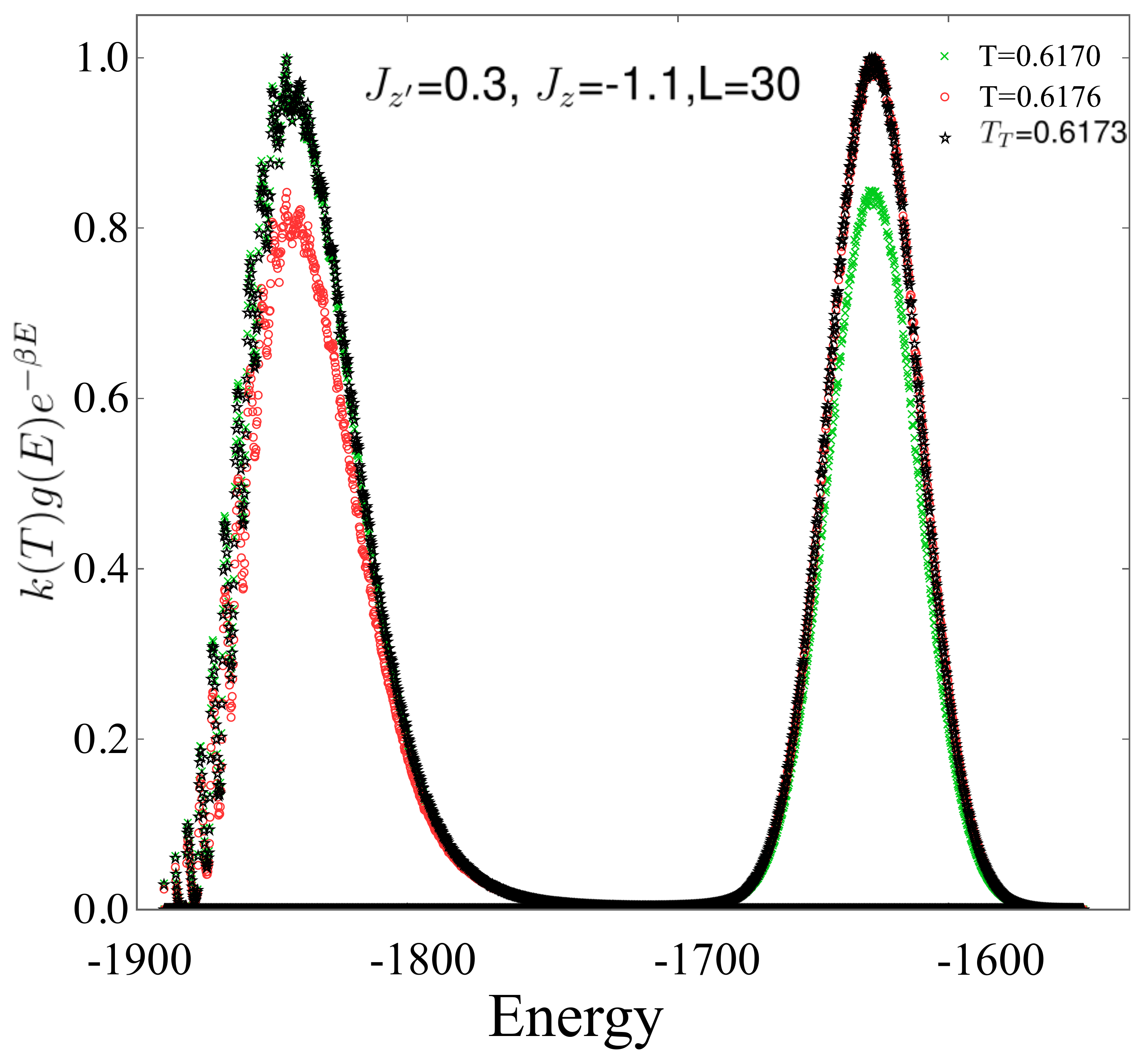}
\caption{(Color online) The canonical distributions for three
temperatures around a first order phase boundary are 
shown (with $J_{z}$ = $-1.1$ and $J_{z^\prime}$=0.3).
The characteristic double peak behavior due to phase coexistence is clear.
At the transition temperature ($T_T$) (black data), the two peaks are of equal height. 
The canonical distributions were normalized by the constant k(T) such that the 
maximum height of the distribution is equal to 1. This simulation was performed on
a size $L$=30 lattice.}
\label{canonicaldistrib}
\end{figure}

This doubly peaked canonical distribution
was found for our model as is shown in Fig.~\ref{canonicaldistrib},
providing additional
confirmation for the existence of the first order phase transition.
For $J_{z} = -1.1$
and $J_{z^\prime}$=0.3, the peaks were found to be of equal height at 
$T_C =$ 0.6173(2). The Metropolis MC data
with the same parameters gave $T_C =$ 0.615(5), which envelopes 
the Wang-Landau value. This procedure confirmed all three 
first order phase transition data points ($J_{z}$=$-1.3,-1.1,-1.0$)
in the $J_{z'}$=0.3 MC phase diagram.

A clear and comprehensive detailing of the WLS
algorithm can be found in the literature
\cite{wang01a,wang01b,landau02}. However, a few specific details of our
simulations are worthy of mention. The energies in our Wang-Landau 
simulation were not binned. In other words,
every unique configuration energy has it's own data point in the density of 
states. Also, windows were not used in the sampling. The entire energy spectrum
shown was sampled in one simulation. Every 10,000 $\times$ 2$N$ spin flips, the
histogram is checked for flatness.  The flatness criterion used is that
no individual energy is visited less than 80 percent of the average
number of visits over all energies. When this criterion is achieved,
the modification factor, f, which was initialized to f=e, is reduced
($f_{i+1} = \sqrt{f_{i}}$), the histogram H(E) is reset to zero and the
process of spin flipping is continued. This algorithm continued until f
was less than $e^{10^{-6}}$ at which time the density of states
converged to our desired level of accuracy. The Wang Landau algorithm
calculates the relative density of states and, therefore, the density of
states was normalized as follows,
\begin{multline}
\ln({g_{\text{normalized}}(E_{i})})=\ln({g_{\text{unnormalized}}(E_{i})})
\\
- \ln({g_{\text{unnormalized}}(E_{GS})}) + \ln({g_{\text{normalized}}(E_{GS})})
\end{multline} 
For the CICM, there are two ground states due to it's spin inversion symmetry. 

\section{Critical exponents}

\begin{figure}[h!]
\includegraphics[height=9.0cm,width=9.0cm]{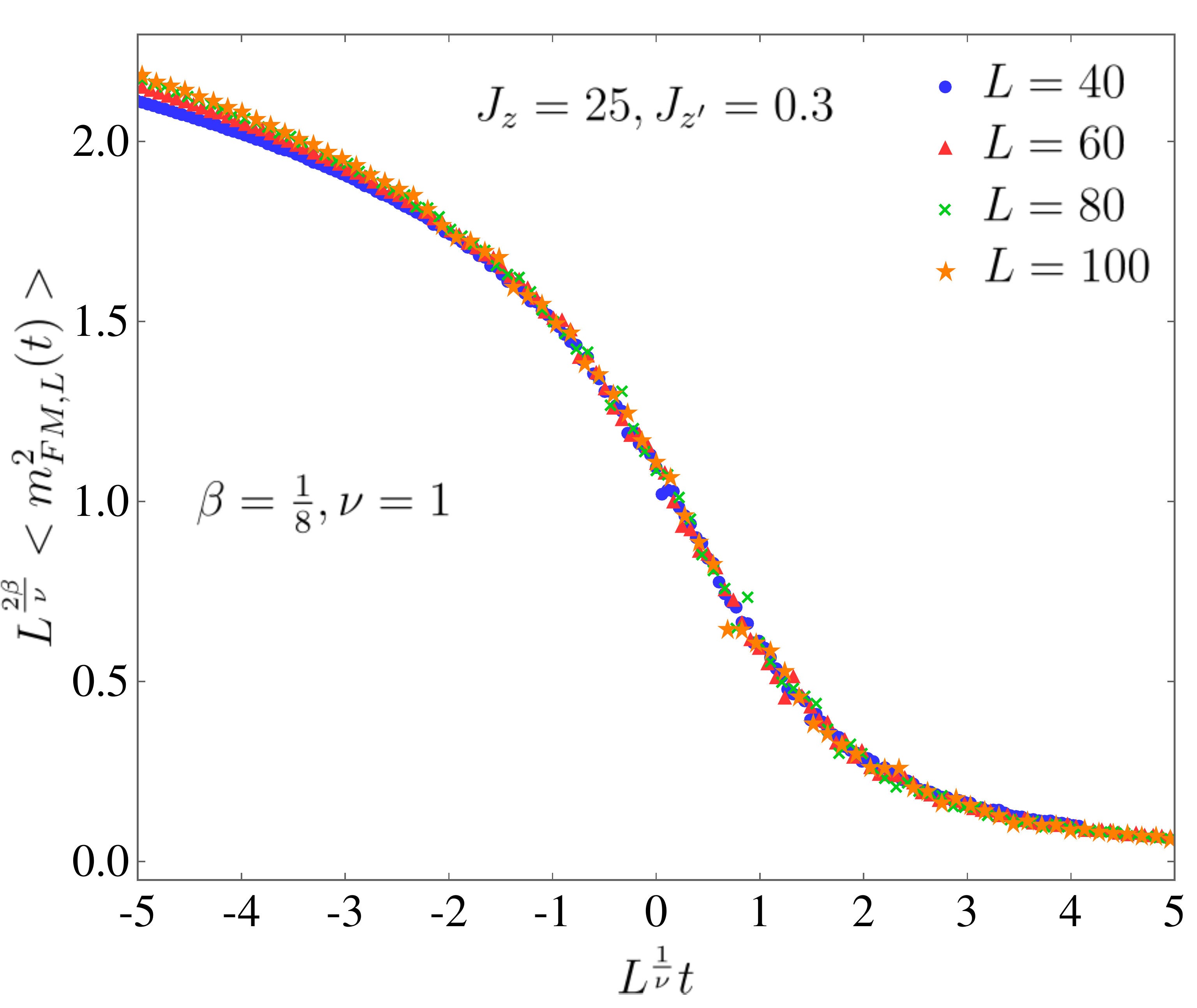}
\caption{(Color online) The data collapse of  ($L^{\frac{2\beta}{\nu}}$ $<m^{2}_{FM}(t)>$) vs. ($L^{\frac{1}{\nu}}$t) for $L$ = 40, 60, 80, and 100 using the two-dimensional Ising universality class exponents is shown.}
\label{m2scaling}
\end{figure}

\begin{figure}
\includegraphics[height=9.0cm,width=9.0cm]{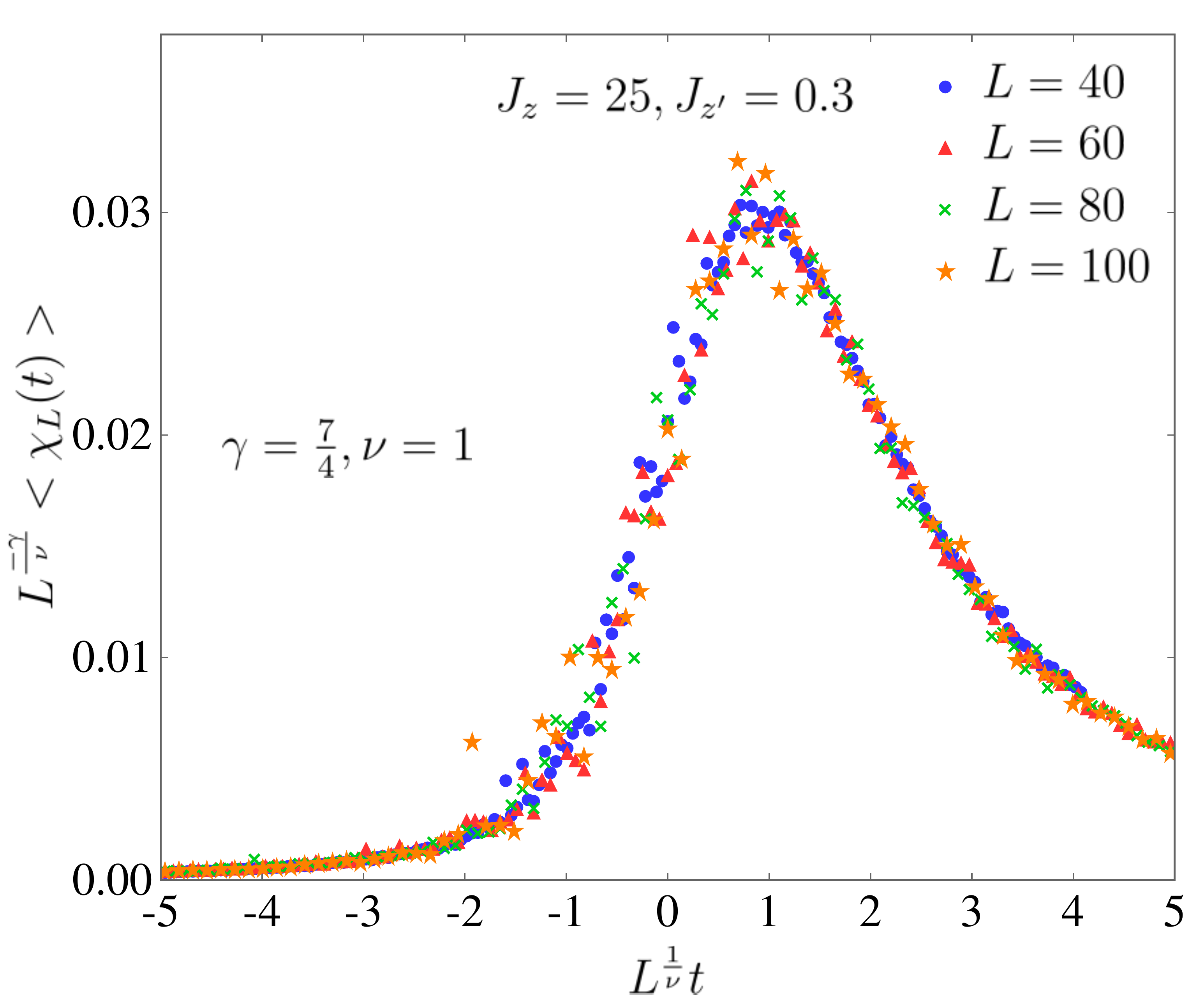}
\caption{(Color online) The data collapse of  ($L^{\frac{-\gamma}{\nu}}$ $<\chi_L(t)>$) vs. ($L^{\frac{1}{\nu}}$t) for $L$ = 40, 60, 80, and 100 using the two-dimensional Ising universality class exponents is shown.}
\label{chiscaling}
\end{figure}

The CICM consists of Ising spins on one dimensional chains with
interchain couplings that connect the system into a two-dimensional lattice and therefore, 
we expect it to belong to the two-dimensional Ising universality class where
the magnetization critical exponent $\beta=\frac{1}{8}$, the correlation length critical exponent 
$\nu = 1$, and the magnetic susceptibility critical exponent $\gamma=\frac{7}{4}$.
To verify this universality class for the CICM (away from the tricritical point) a finite size scaling analysis was performed. Plots of ($L^{\frac{2\beta}{\nu}}$ $<m^{2}_{FM}(t)>$) vs. ($L^{\frac{1}{\nu}}$t) and ($L^{\frac{-\gamma}{\nu}}$ $<\chi_L(t)>$) vs. ($L^{\frac{1}{\nu}}$t) for various values of $L$ will collapse onto a single curve for the correct values of the exponents $\beta$, $\nu$, and $\gamma$ \cite{newman99}.
We measured $<m_{FM}^{2}>$ and $<\chi_L>$ as a function of temperature for the CICM with $J_z$=25 and $J_{z'}$=0.3. For these parameters, $T_C$ = 3.63. This was used to define the reduced temperature t = $\frac{T-T_C}{T_C}$. Fig.~\ref{m2scaling} and Fig.~\ref{chiscaling} show the results of this analysis. The data collapses nicely over a broad range of temperatures. This provides a satisfying consistency check to our expectation of the universality class of the CICM.

Precisely at a tricritical point,
the critical exponents are known to take on different tricritical values \cite{landau00}.
We attempted to measure the tricritical exponents at the tricritical point in
our model but this proved to require a level of precision beyond the scope of our work.
However, we did find that when applying the same finite size scaling analysis that is detailed 
in the previous paragraph, including using the same two-dimensional Ising exponents, to a tricritical
point in our model ($J_z$=-0.9 and $J_{z'}$=0.3), there was a significant decrease in the
degree to which the data ``collapsed." Although inconclusive, this finding is consistent with 
our expectation that there will be a change in exponents at the tricritical point. 

Finally, the critical Binder cumulant, U* is the value of the Binder cumulant at the critical temperature in the thermodynamic limit. For the 2D square Ising model it has been shown that U*=0.61069... \cite{Kamieniarz}. For $J_z$=25 and $J_{z'}$=0.3 in the CICM, our data shows that U* is somewhere between 0.605 and 0.615,  consistent with the known value for 2D Ising universality. At $J_z$=-0.9 and $J_{z'}$=0.3 (approximate location of a tricritical point) our data has a larger spread of possible U* values although it is clearly less than 0.61069. U* at the tricritical point is in the range 0.50 to 0.55. We also measured U* at the tricritical point of the 2D Blume-Capel model and a similar range of values was found, providing some evidence in favor of 2D tricritical Ising universality. 

\section{Conclusions}

Using a combination of mean field theory, Metropolis MC, and
Wang-Landau simulations, we have explored an Ising-like model on a
lattice composed of a 1D$\times$1D collection of coupled chains. As is well
known, 1D Ising chains with short range interactions
do not exhibit finite temperature ordered phases.
However, interchain couplings connect the chains into a 2D framework
which shows multiple ordered phases at finite temperatures. The phase
transitions between the ordered and disordered phases can be of first or
second order as evidenced by the behavior of the Binder fourth order
cumulants and the canonical distributions. The existence of tricritical
points in the phase diagram depends on the value of
$J_{z^\prime}$. According to the MC simulations, for $J_{z^\prime}$=0.1
and 1.0, there are no tricritical points but for intermediate $J_{z^\prime}$=0.3, there are
tricritical points.

It would be interesting to see if the Nb$_{12}$O$_{29}$ materials can
be tuned between first and second order transitions by varying
pressure, doping or other parameters, thus giving rise to novel
realizations of tricritical systems.

In some materials which exhibit this 1D$\times$1D geometry, the quantum mechanical
nature of the degrees of freedom may be crucial to the observed
phenomena.  For example, in the optical lattice case, the
focus is on the occurrence of Bose-Einstein condensation
at finite momentum, and in a pattern of
orbitals which alternates as $p_x \pm i p_y$ on the two sublattices.
Our work shows that even at the classical level, these
crossed-chains systems exhibit complex phase-transitions
and crossovers. Future work could address the additional
non-trivial physics which arises when the phase transitions
are driven to $T=0$, giving rise to exotic quantum phase transitions.
Additional future work could study the critical and tricritical exponents of this model
with greater precision and breadth. 

\paragraph{}

TC and RTS were supported by Department of Energy grant DE-SC0014671.
The work of RRPS is supported by the US National Science
Foundation grant number DMR-1306048.


\end{document}